\documentclass[conference]{IEEEtran}
\IEEEoverridecommandlockouts

\usepackage{cite}
\usepackage{amsmath,amssymb,amsfonts}
\usepackage{algorithmic}
\usepackage{graphicx}
\usepackage{textcomp}
\usepackage{xcolor}
\usepackage{url}
\usepackage{float}

\newcommand{\RN}[1]{%
  \textup{\uppercase\expandafter{\romannumeral#1}}%
}

\def\BibTeX{{\rm B\kern-.05em{\sc i\kern-.025em b}\kern-.08em
    T\kern-.1667em\lower.7ex\hbox{E}\kern-.125emX}}
\begin{document}

\title{Reducing Harmonic Distortion in a 5-Level Cascaded H-bridge Inverter Fed by a 12-Pulse Thyristor Rectifier \\

\thanks{}
}

\author{\IEEEauthorblockN{ Milad Sadoughi}
\IEEEauthorblockA{\textit{Faculty of Electrical,} \\
\textit{and Computer Engineering}\\
\textit{Urmia University}\\
			Urmia, Iran \\
\url{st_m.sadughi@urmia.ac.ir}}
\and
\IEEEauthorblockN{Amirhossein Pourdadashnia}
\IEEEauthorblockA{\textit{Faculty of Electrical,} \\
\textit{and Computer Engineering}\\
\textit{Urmia University}\\
			Urmia, Iran \\
\url{st_a.pourdadashnia@urmia.ac.ir}}
\and
\IEEEauthorblockN{Mohammad Farhadi-Kangarlu}
\IEEEauthorblockA{\textit{Faculty of Electrical,} \\
\textit{and Computer Engineering}\\
\textit{Urmia University}\\
			Urmia, Iran \\
\url{m.farhadi@urmia.ac.ir}}
\and
\IEEEauthorblockN{Sadjad Galvani}
\IEEEauthorblockA{\textit{Faculty of Electrical,} \\
\textit{and Computer Engineering}\\
\textit{Urmia University}\\
			Urmia, Iran \\
\url{s.galvani@urmia.ac.ir}}

}


\maketitle

\begin{abstract}
This paper investigates a multilevel inverter with a capability that produces a wide voltage range with high quality. The selective harmonic elimination (SHE) method is considered for a single-phase 5-level cascaded H-bridge (CHB) inverter, in which the particle swarm optimization (PSO) algorithm solves the nonlinear equations. However, eliminating the low-order harmonics has been challenging when a low range of output voltage is required. To surmount such challenges and access to a wide range of output voltage, an adjustable dc-link is introduced that allows the inverter to increase the modulation index, resulting in a significant decrease in total harmonic distortion (THD).  In this paper, to regulate the dc-link voltage amount, a 12-pulse rectifier is employed to allow the inverter to produce the output voltage requirements with less distortion. To prove such claims, the PSO algorithm is modified to calculate the optimal angles, as a result, the switching angles are applied in SIMULINK MATLAB to generate a 5-level output voltage.
\end{abstract}
\vspace{0.5cm}

\section{Introduction}
Multilevel inverters (MLIs) have captured industrial's attention due to their capability to produce high-quality voltage. They possess some advantages that distinguish them from two-level inverters. Such benefits include reducing voltage stress on power electronics semiconductors, reducing total harmonic distortion (THD), lower electromagnetic interference (EMI), lower switching losses, etc. \cite{1,2}. In resilient improvement, the PV-based inverters play an important role; more information regarding the smart PV inverters can be referred to \cite{19,20}.

There has been studied various topologies of MLIs in many pieces of literature. The cascaded H-bridge (CHB) inverter has achieved more popularity than other structures as they have high modularity and fault tolerance \cite{3,4}. This inverter consists of a series of combination of H-bridges that provides the inverter to increase its voltage levels. Besides, their isolated dc sources can give them a possibility to behave symmetrically if the dc sources are equal. In symmetric mode, their voltage levels are restricted to $2N+1$, in which $N$ is the number of isolated dc sources. Hence, by increasing the number of bridges, the inverter will be able to produce high-quality voltage.

In recent years, several switching methods have been carried out for the MLIs. Pulse width modulation (PWM) is widely used in industry, which consists of carrier-based PWM (C-PWM), space vector PWM (SVPWM), and selective harmonic elimination PWM \cite{18,5,6,7}. The SHE-PWM method is based on the SHE method that has facilitated many aspects of switching since such a technique operates at low frequency and hence, leads the switching losses to decrease \cite{8,9}. In the SHE technique, the low-order harmonics are eliminated by a series of nonlinear equations, which causes the inverter to have a filter with a small size in its output. The number of nonlinear equations is determined regarding the number of dc sources. In this paper, a single-phase 5-level CHB inverter has been examined with two separated dc sourced. Consequently, there would be two equations to calculate the optimal angles. As the system is regarded as a single-phase, the tripled harmonic should be considered; hence, the 3\textsuperscript{rd} harmonic is considered to eliminate. In the SHE technique, solving the nonlinear equations are extremely complicated, and such complexity will be raised if the number of equations increases. To deal with the aforementioned challenge, several evolutionary optimization techniques can be utilized, such as particle swarm optimization (PSO), teaching learning-based algorithm (TLBO), Artificial bee colony (ABC) algorithm, genetic algorithm (GA), etc. \cite{10,11}. One of the most powerful and most expeditious methods based on evolutionary algorithms is recommended in this research. Particle swarm optimization (PSO) is a potent algorithm for solving the optimal angles required by the SHE technique equations, which can quickly solve the equations with high accuracy.

Producing a high-quality voltage by the MLI is enormously beneficial in many circumstances, such as drive applications and voltage compensators, where the quality of the output voltage is critical \cite{12,13}. In the SHE method, despite the low-order harmonics being eliminated efficiently, the THD amount is remarkably high in low ranges of the modulation index \cite{16}. To surmount such a problem, this paper has expressed a proposed method by which the inverter can generate a high-quality voltage in a wide range and has compared the suggested technique with previous methods. Thereby, section \RN{2} explains the SHE method's background, and the evolutionary algorithms are introduced to solve the nonlinear equations. In section \RN{3}, the proposed method is introduced to overcome the problems are arising in section \RN{2}, and accordingly, the simulation results are reported in section \RN{4}. Eventually, to summarize the work, the conclusions are asserted in section \RN{5}.

\section{Downsides of the SHE Method}
Fig. 1 (a) illustrates a single-phase 5-level CHB inverter composed of two separate dc sources. Due to this fact, the output voltage waveform is demonstrated in Fig. 1 (b), in which $\theta_1$ and $\theta_2$ are considered as switching angles. In the SHE technique, the Fourier series expansion of the output voltage waveform is considered as
\begin{equation}\small
{V_o}(t) = \frac{{4{V_{dc}}}}{\pi }\sum\limits_{n = 1, 3, 5, ...}^\infty  {\frac{1}{n}\left\{ {\cos (n{\theta _1}) + \cos (n{\theta _2})} \right\}} \sin (n\omega t)
\end{equation}
subject to
\begin{equation}\small
0 \leqslant {\theta _1} < {\theta _2} \leqslant 90
\end{equation}
To calculate the switching angles, the SHE equations should be solved by evolutionary algorithms. However, solving such equations is most challenging. Accordingly, the SHE equations are considered as follows:
\begin{equation}\small
{({V_o})_1} = \frac{{4{V_{dc}}}}{\pi }\left( {\cos ({\theta _1}) + \cos ({\theta _2})} \right)
\end{equation}
\begin{equation}\small
{({V_o})_3} = \frac{{4{V_{dc}}}}{\pi }\left( {\cos (3{\theta _1}) + \cos (3{\theta _2})} \right)
\end{equation}
The first equation is solved to adjust the inverter output voltage equal to the desired AC voltage, and the second one is solved to eliminate the 3\textsuperscript{rd} harmonic, as the system is supposed to be a single-phase \cite{15}.
 
As mentioned before, the SHE equations are highly challenging to calculate the switching angles. To surmount this problem, various algorithms have been introduced, and in this paper, the optimal angles are obtained by the PSO algorithm. A cost function is defined in such an algorithm to set the desired amount's output voltage and eliminate the low-order harmonics. The cost function is defined as follows
\begin{equation}\small
f({\theta _1},{\theta _2}) = \left| {M - \frac{{\left| {{V_1}} \right|}}{{N{V_{dc}}}}} \right| + \frac{{\left| {{V_3}} \right|}}{{N{V_{dc}}}}
\end{equation}
where $M$ is the modulation index, and $N$ is the number of isolated dc sources. The modulation index is the ratio of the fundamental output voltage to the total of the dc-links
\begin{equation}\small
{M_{old}} = {V_{o, pu}} = \frac{{{{({V_o})}_1}}}{{2{V_{dc}}}}
\end{equation}
where ($V_{o, pu}$) is the per-unit of the inverter output voltage.       

By applying the switching angles, the simulation results are achieved, which is demonstrated in Fig. 2. Accordingly, in the low ranges of the modulation index, the THD amount is exceptionally high, and therefore, a filter with significant components is required \cite{14,17}.
\begin{figure}[]
       \includegraphics[width=8.8cm]{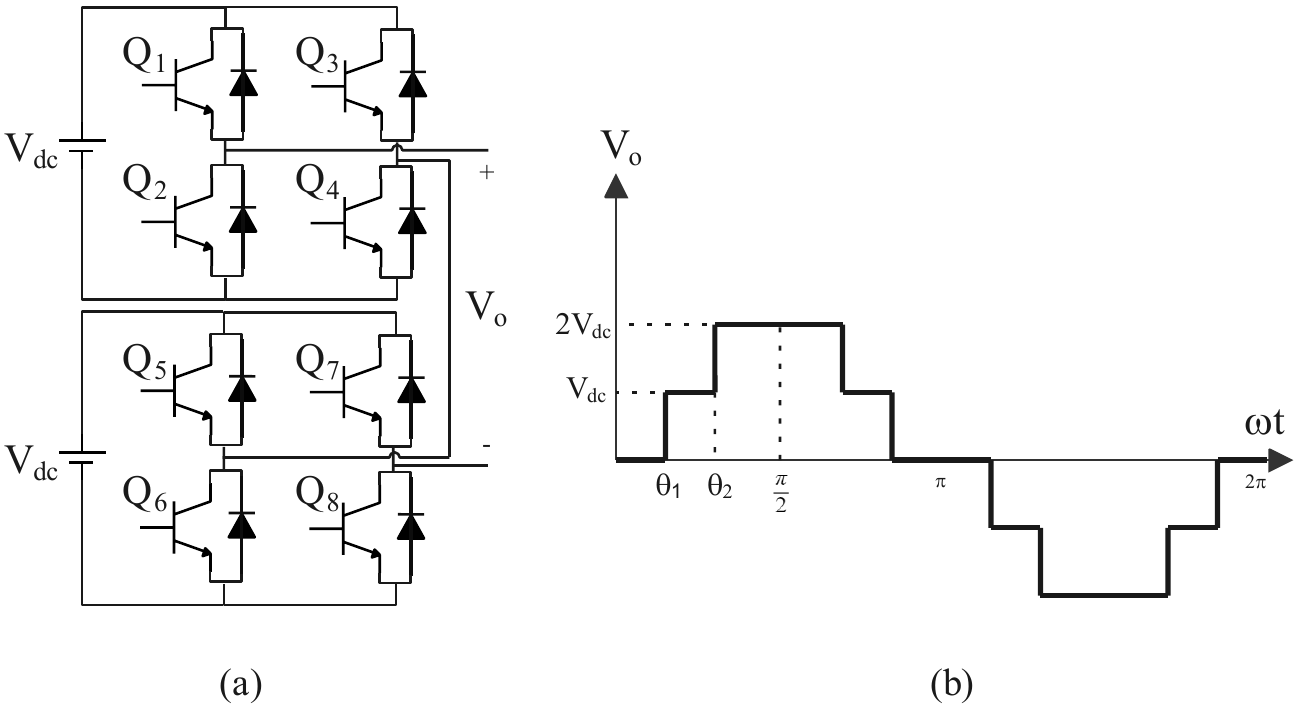}
	   \caption{(a) A 5-level CHB inverter. (b) The output voltage waveform using the SHE technique.}
       \label{Fig}
\end{figure}

\section{Proposed Method}
In the previous section, the optimal angles were obtained by the PSO algorithm, and then the simulation-based results were shown in Fig. 2. In this section, to overcome the problems of the SHE method, a variable dc-link is proposed. By regulating the dc-link voltage, the modulation index can be increased, and thus, the harmonic distortion is reduced significantly. Hence, when a voltage at low ranges is required, using the proposed method, the inverter produces such a voltage with high quality. 

The proposed scheme is depicted in Fig. 3, in which the inverter consists of two bridges with isolated dc-links. To variate the amount of the dc-links voltage, a twelve-pulse active rectifier is considered, wherein a phase-shifting transformer powers two six-pulse thyristor rectifiers with two secondary winding in star and delta connections, and consequently, both dc-links could be regulated in a way that the inverter maintains in symmetric mode. Also, the active rectifier is controlled to ensure that the modulation index is increased equal to 1, whereby the inverter is capable of maintaining the THD in minimum value. 
\begin{figure}[]
       \includegraphics[width=9cm]{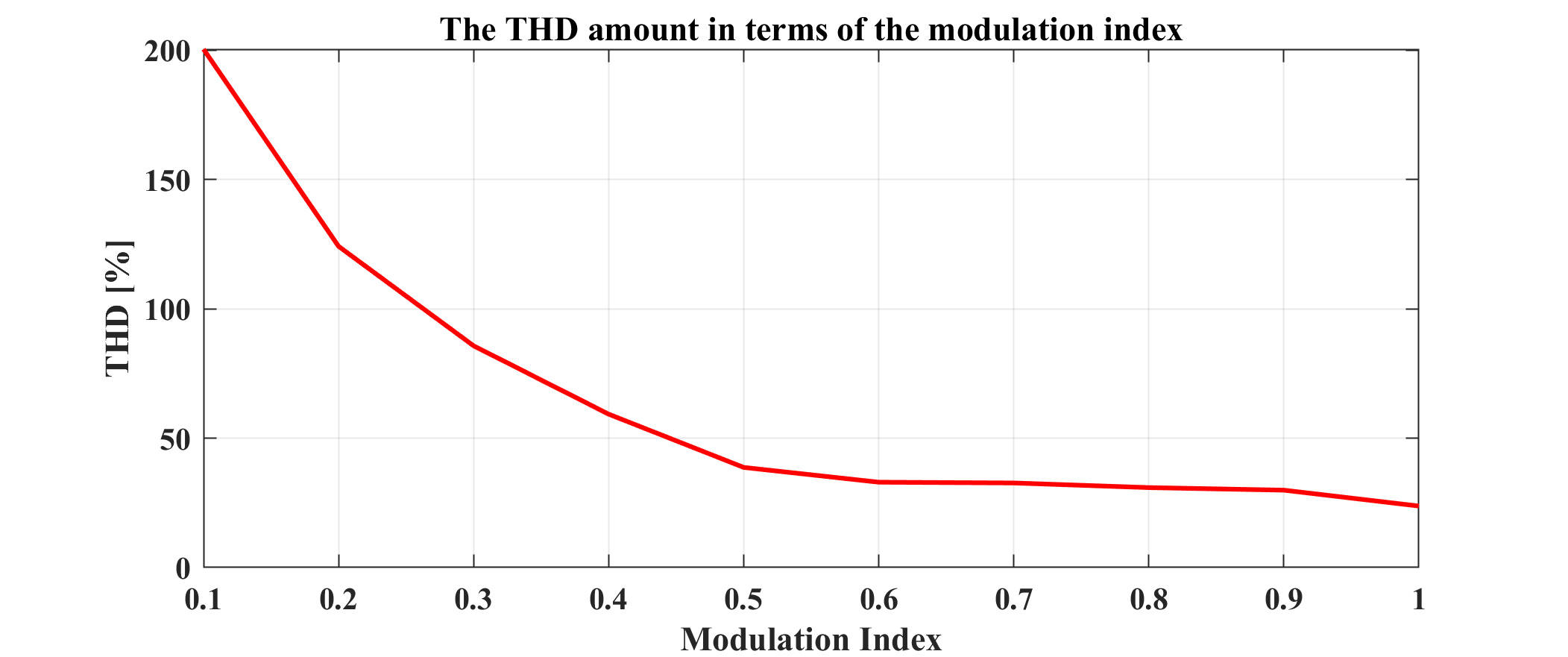}
	   \caption{Variations of the THD in various ranges of the modulation index.}
       \label{Fig}
\end{figure}

In the controlled rectifier, the dc-link voltage is obtained as follows:
\begin{equation}\small
E = \frac{3}{\pi }{V_s}\cos (\alpha )
\end{equation}
where $E$ is the produced voltage by the rectifier, $V_s$ is the line-to-line voltage on the secondary side of the transformer, and $\alpha$ is the firing angle. To enhance the modulation index equal to 1, the following equation should be considered.
\begin{equation}\small
{M_{new}} = \frac{{{{({V_o})}_1}}}{{2E}}\xrightarrow{{{M_{new}} = 1}}{({V_o})_1} = 2E
\end{equation}
In the proposed method, $(V_o)_1$ is the desired output voltage that is the same value in (6), hence, according to (6) and (8), the regulated dc-link is adjusted in the following way
\begin{equation}\small
\left. \begin{gathered}
  {({V_o})_1} = 2E \hfill \\
  {M_{old}} = {V_{o, pu}} = \frac{{{{({V_o})}_1}}}{{2{V_{dc}}}} \hfill \\ 
\end{gathered}  \right\} \Rightarrow E = {V_{o, pu}}{V_{dc}}\end{equation}
where $V_{dc}$ is the isolated dc source demonstrated in Fig. 1. Eventually, using (7) and (9), the rectifier's firing angle is obtained in order to regulate the dc-link voltage in a way that the value of the modulation index is equal to 1.
\begin{equation}\small{V_{o, pu}}{V_{dc}} = \frac{3}{\pi }{V_s}\cos (\alpha )   \Rightarrow   \alpha  = {\cos ^{ - 1}}(\frac{{\pi {V_{o, pu}}{V_{dc}}}}{{3{V_s}}})\end{equation}
Then, based on the modulation index ($M=1$) and the dc-link value, the PSO algorithm recalculated the optimal angles. Accordingly, the simulation-based results are achieved that mentioned in the next section.

\section{Simulation Results}
To verify the proposed method’s result, the studied system has simulated in MATLAB SIMULINK, and results are shown to prove the significant improvement in the THD amount. As illustrated in Fig. 1 (a), the system composed of two separated dc sources, and each of them is considered 200 volts. For the base-case, the simulation has been carried out for all modulation index ranges, and Fig. 2 provides such results; hence, in high ranges (i.e., above 0.8), the amount of the THD is not extremely high, and based on this fact, the low range is examined by detail. Nevertheless, the output voltage distortion value is examined for all ranges and is provided in Table \RN{1}.
\begin{figure}[]
       \includegraphics[width=8.5cm]{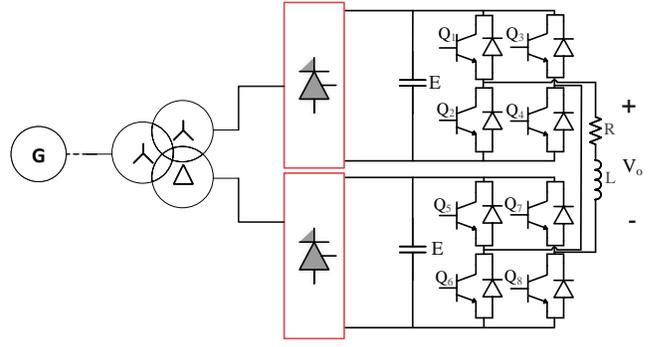}
	   \caption{A twelve-pulse thyristor rectifier to fed the 5-level CHB inverter using the proposed method.}
       \label{Fig}
\end{figure}
The switching angles are the principal element in producing the waveform and also in its quality. These angles tend to be in their maximum value by reducing the modulation index. In the proposed method, increasing the modulation index to 1 allows the PSO algorithm to calculate the angles in such a way that the waveform maintains in a high-quality manner, resulting in a minimum distortion in the output voltage. The switching angles obtained by the PSO algorithm for both methods are shown in Fig. 4, in which the influences as mentioned earlier are shown distinctly. Finally, such angles are stored in a look-up table (LUT), and therefore, the control system can recognize the required voltage and allow the inverter to set the output voltage with minimum distortion.
\begin{figure}[]
\centering
\begin{tabular}{cc}
 \includegraphics[width=8cm]{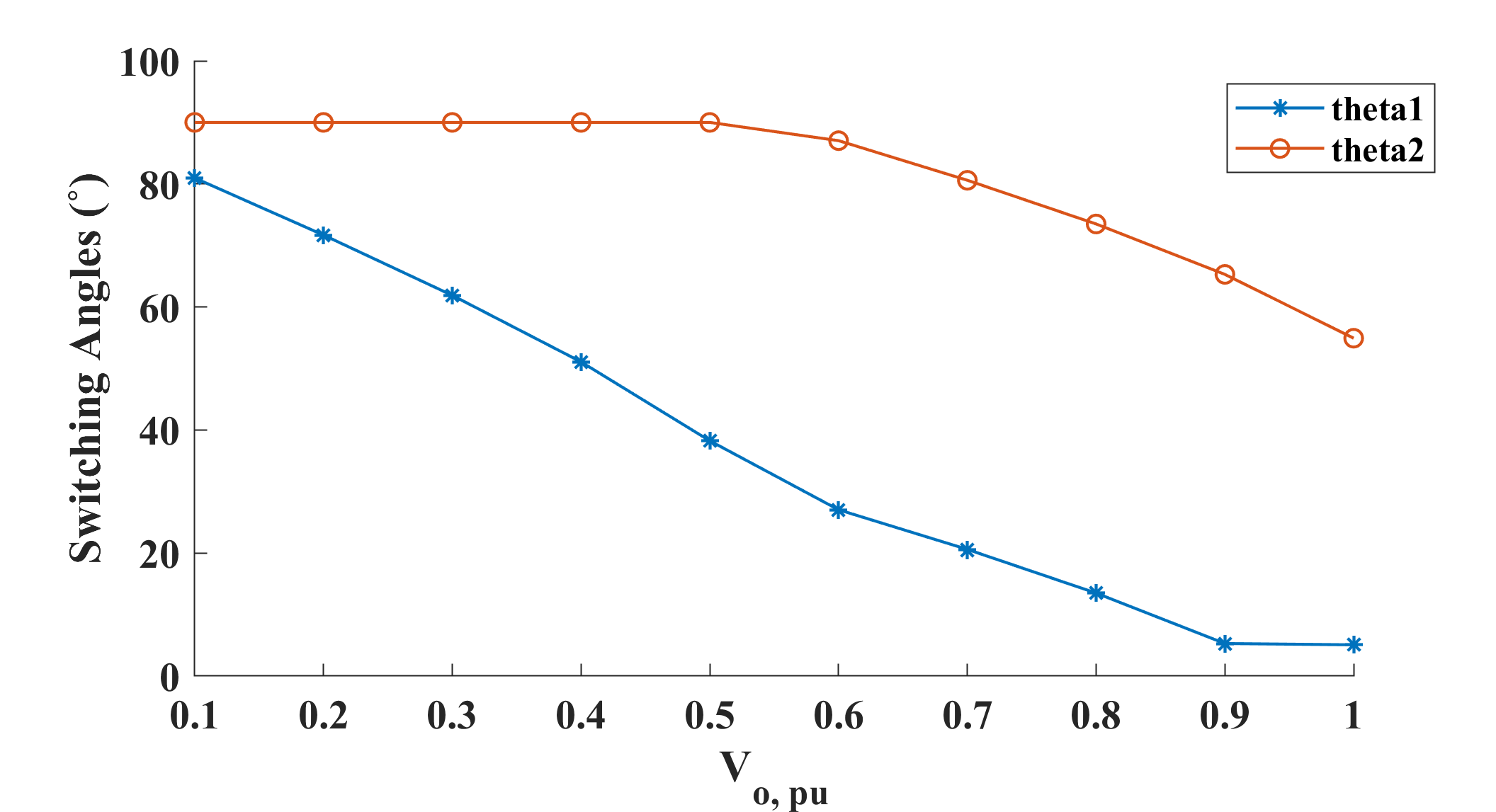}\\ 
\footnotesize (a)\\
\includegraphics[width=8cm]{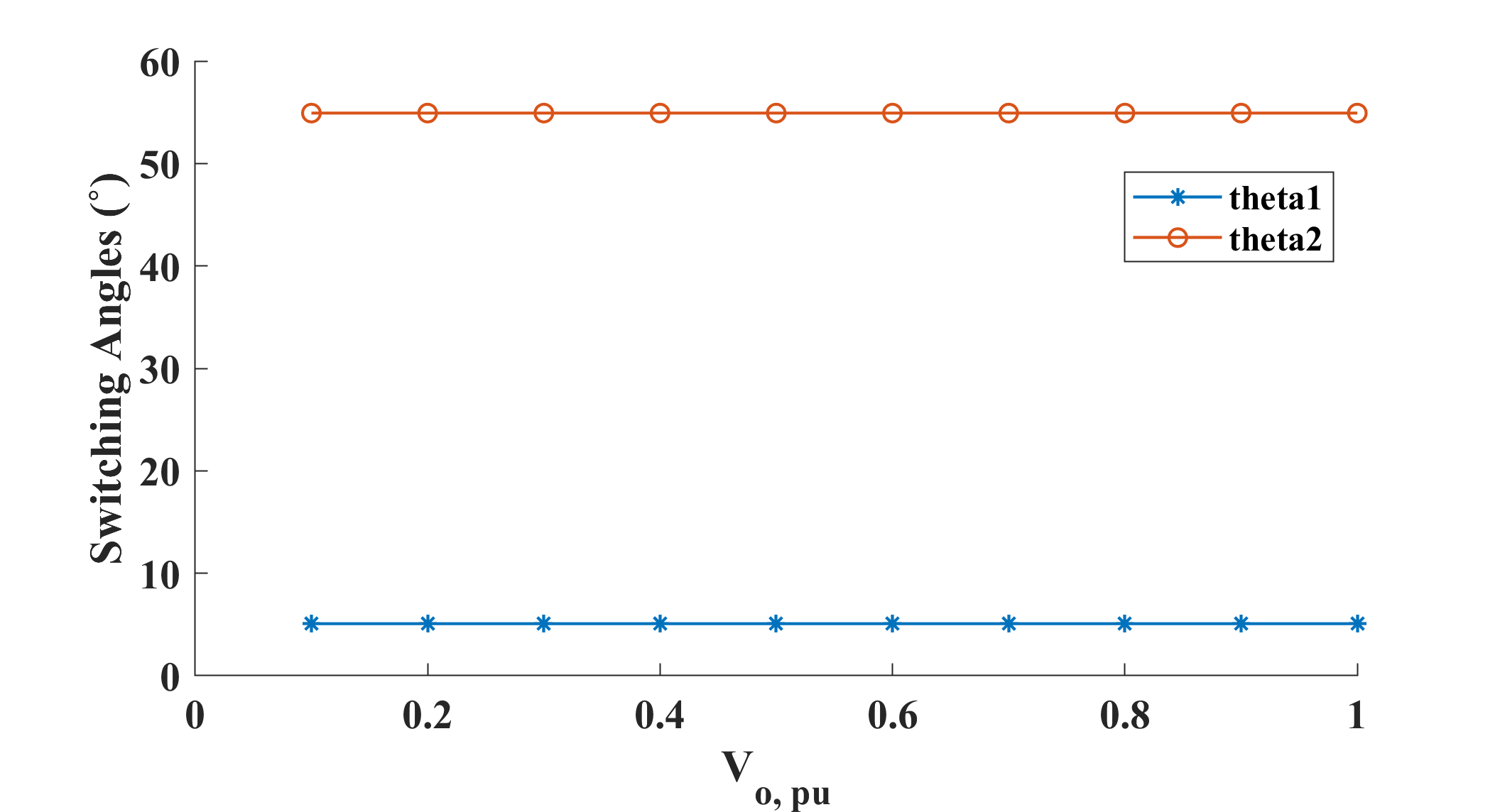} \\
\footnotesize(b)
\end{tabular}
\caption{The switching angles versus the modulation index for a 5-level CHB inverter. (a) Conventional Method. (b) Proposed Method.}
\label{Fig.}
\end{figure}

\begin{figure}[]
\centering
\begin{tabular}{cc}
 \includegraphics[width=8cm]{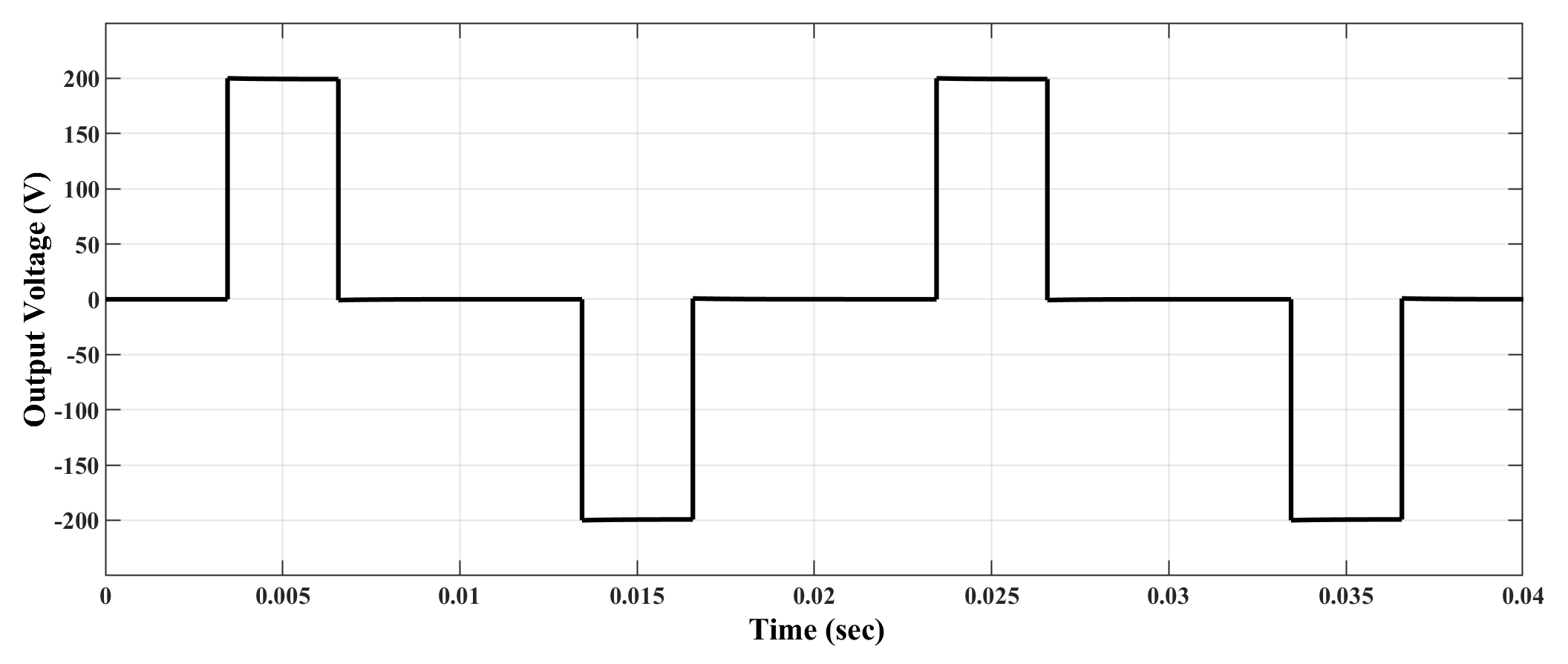}\\ 
\footnotesize (a)\\
\includegraphics[width=8cm]{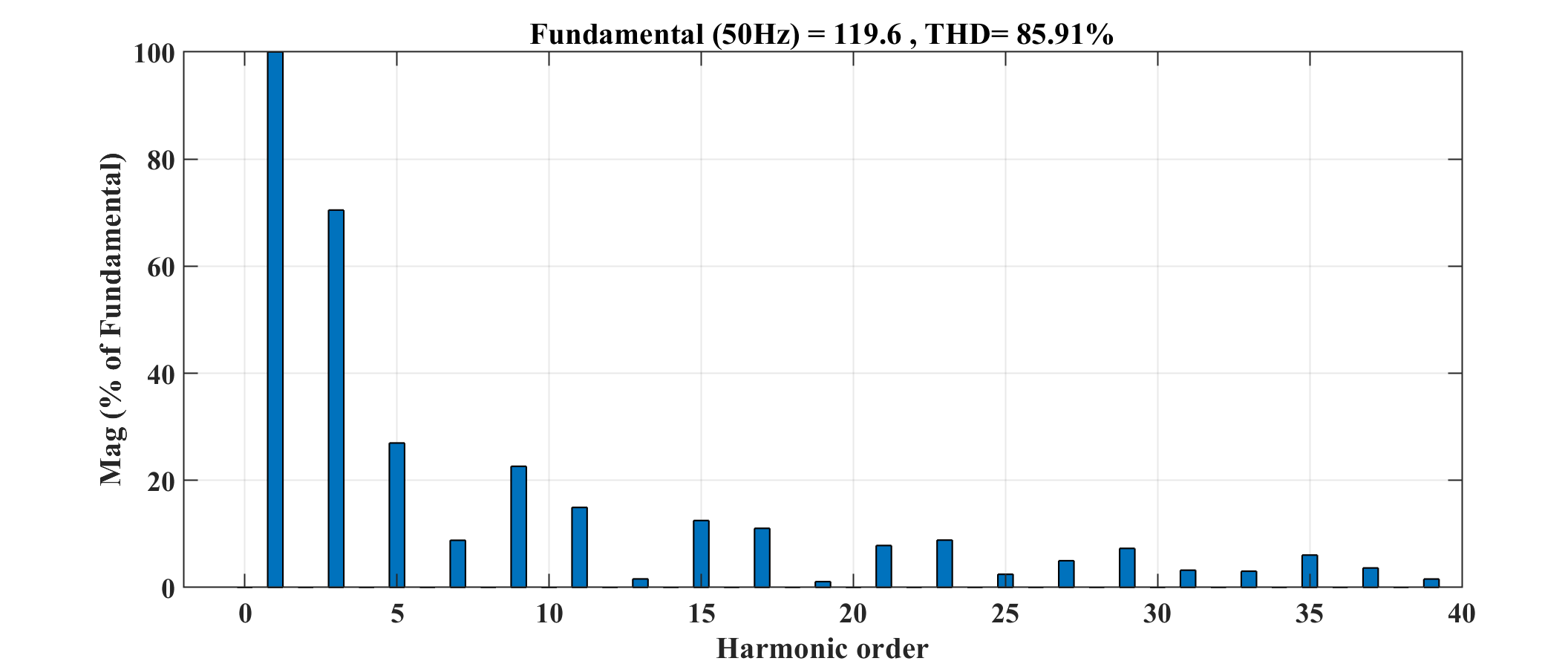} \\
\footnotesize(b)
\end{tabular}
\caption{The base-case system with $M_{old}$=$V_{o, pu}$=0.3. (a) Output voltage waveform. (b) FFT analyze.}
\label{Fig.}
\end{figure}
\begin{figure}[]
\centering
\begin{tabular}{cc}
 \includegraphics[width=8cm]{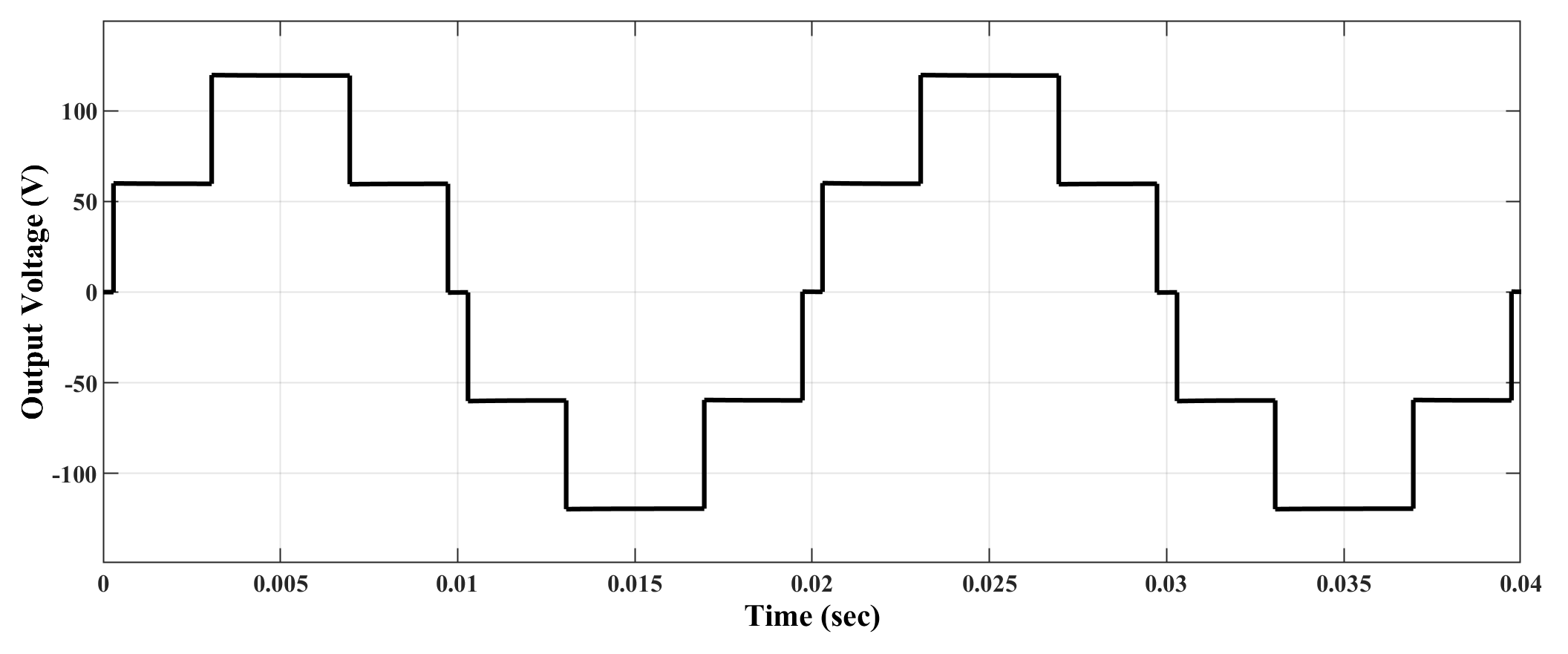}\\ 
\footnotesize (a)\\
\includegraphics[width=8cm]{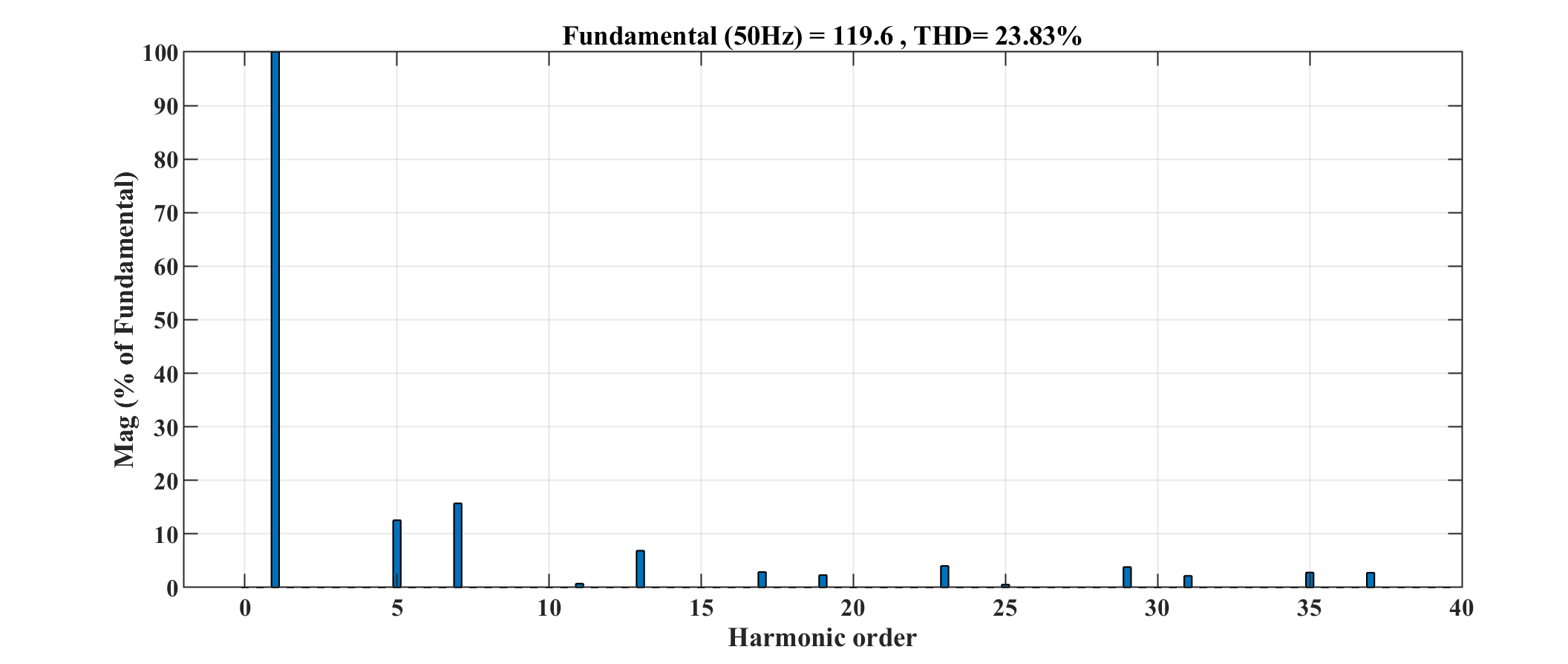} \\
\footnotesize(b)
\end{tabular}
\caption{The proposed system with $V_{o, pu}$=0.3 and $M_{new}$=1. (a) Output voltage waveform. (b) FFT analyze.}
\label{Fig.}
\end{figure}
To investigate the THDs amount and waveforms quality, two samples of the required output voltage are examined for 0.3 and 0.6 per unit, by which the inverter produces 120 and 240 volts, respectively. Fig. 5 shows the simulation results in the base-case of the 5-level CHB inverter, in which the modulation index is equal to 0.3. As shown in Fig. 5(a), the number of the output voltage levels has reduced to three as one of the switching angles $\theta_2$ is equal to 90. Hence, the 3\textsuperscript{rd} harmonic is not eliminated by the PSO algorithm, resulting in a considerable THD increase that is appeared in Fig. 5(b).

On the other hand, to produce the same fundamental output voltage by the 5-level CHB inverter, an active rectifier is employed to adjust both dc-links values. According to (6), the $V_{o, pu}$ is equal to 0.3, and both dc-links voltages should be adjusted on 60 volts to increase the modulation index equal to 1. Thereby, according to the desired voltage, the control system can choose the proper optimal angles from the LUTs, and then the inverter produces the same magnitude output voltage with low distortion. Fig. 6 demonstrates the output voltage waveform and the FFT analysis. The voltage levels are increased to five, and the THD amount is improved significantly. Moreover, by comparing Figs. 5(b) and 6(b), it can be recognized that in the base-case study, in low ranges of $V_{o, pu}$, the 3\textsuperscript{rd} harmonic cannot entirely be removed, however, in the proposed scheme, such a harmonic is eliminated.

In order to investigate 240 volts ($V_{o, pu}$=0.6), the aforementioned process has been done, and the results are carried in Figs. 7 and 8 for the conventional and proposed methods, respectively. Also, the THD improvement has been carried out for other ranges of the modulation index and brought in Table \RN{1}. As reported in this tabular, the THD is fixed in a constant value, whereby the inverter can operate in a way that the $M$ is equal to 1.

\begin{figure}[]
\centering
\begin{tabular}{cc}
 \includegraphics[width=8cm]{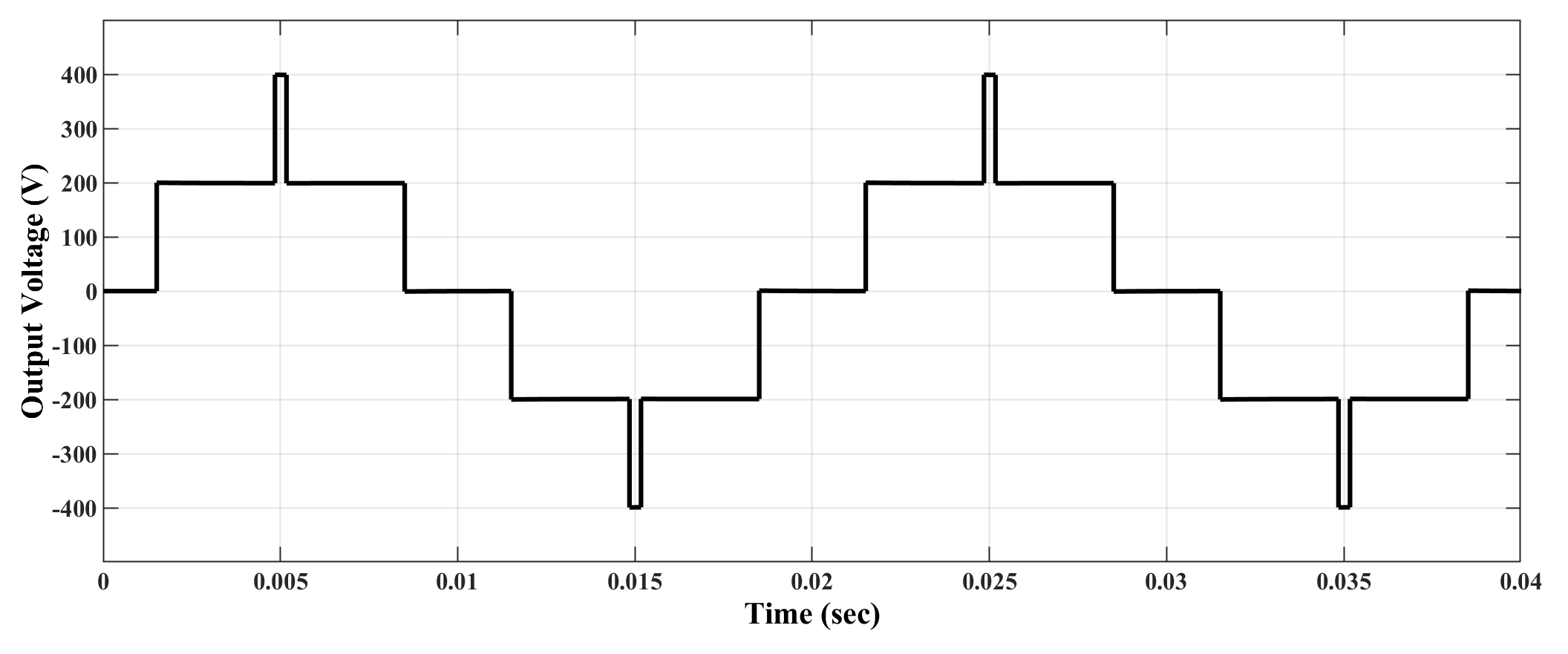}\\ 
\footnotesize (a)\\
\includegraphics[width=8cm]{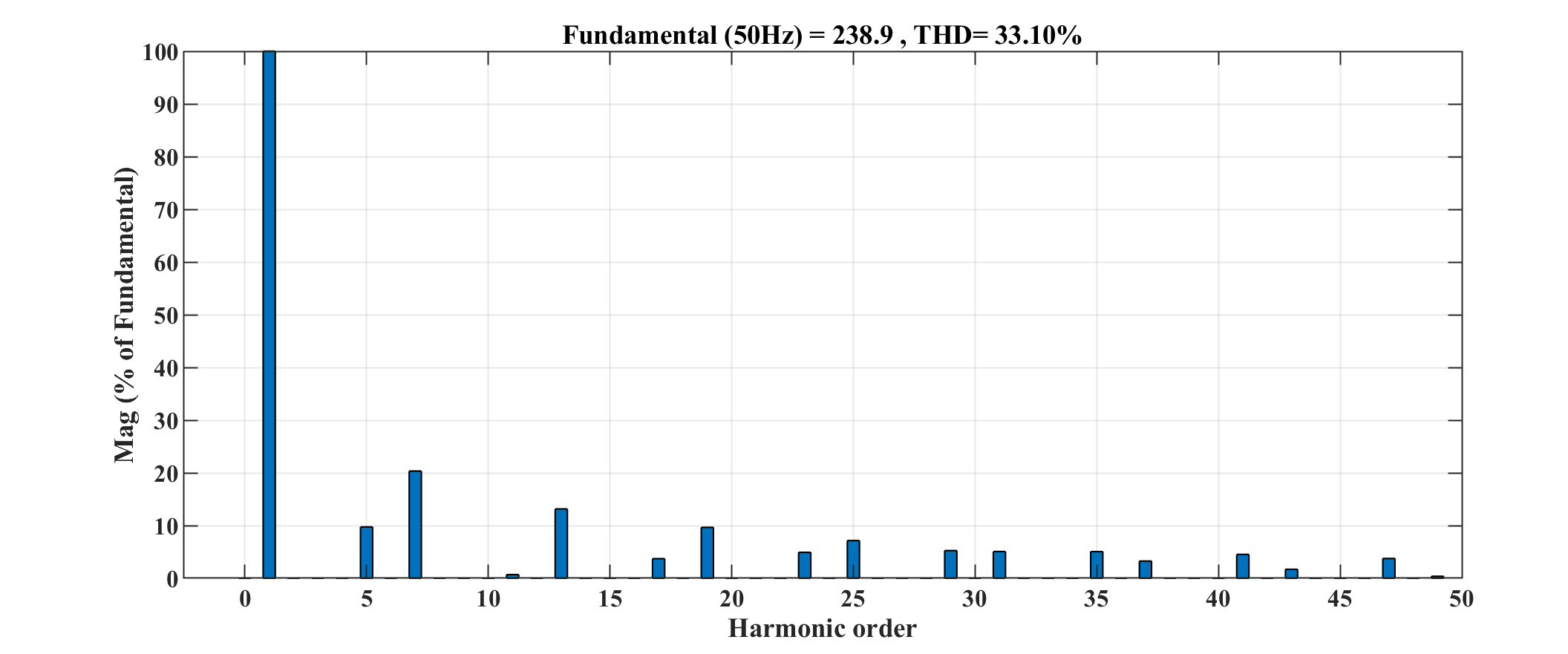} \\
\footnotesize(b)
\end{tabular}
\caption{The base-case system with $M_{old}$=$V_{o, pu}$=0.6. (a) Output voltage waveform. (b) FFT analyze.}
\label{Fig.}
\end{figure} 
\begin{figure}[]
\centering
\begin{tabular}{cc}
 \includegraphics[width=8cm]{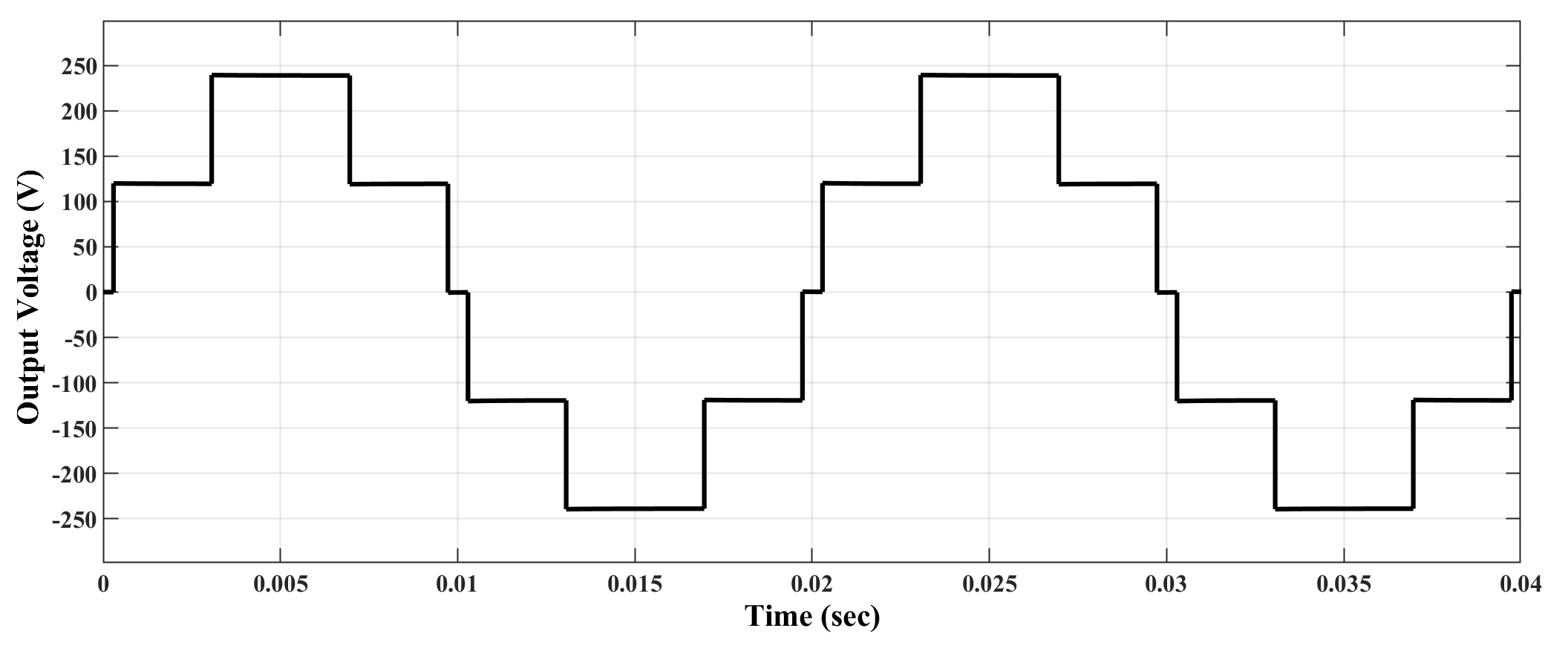}\\ 
\footnotesize (a)\\
\includegraphics[width=8cm]{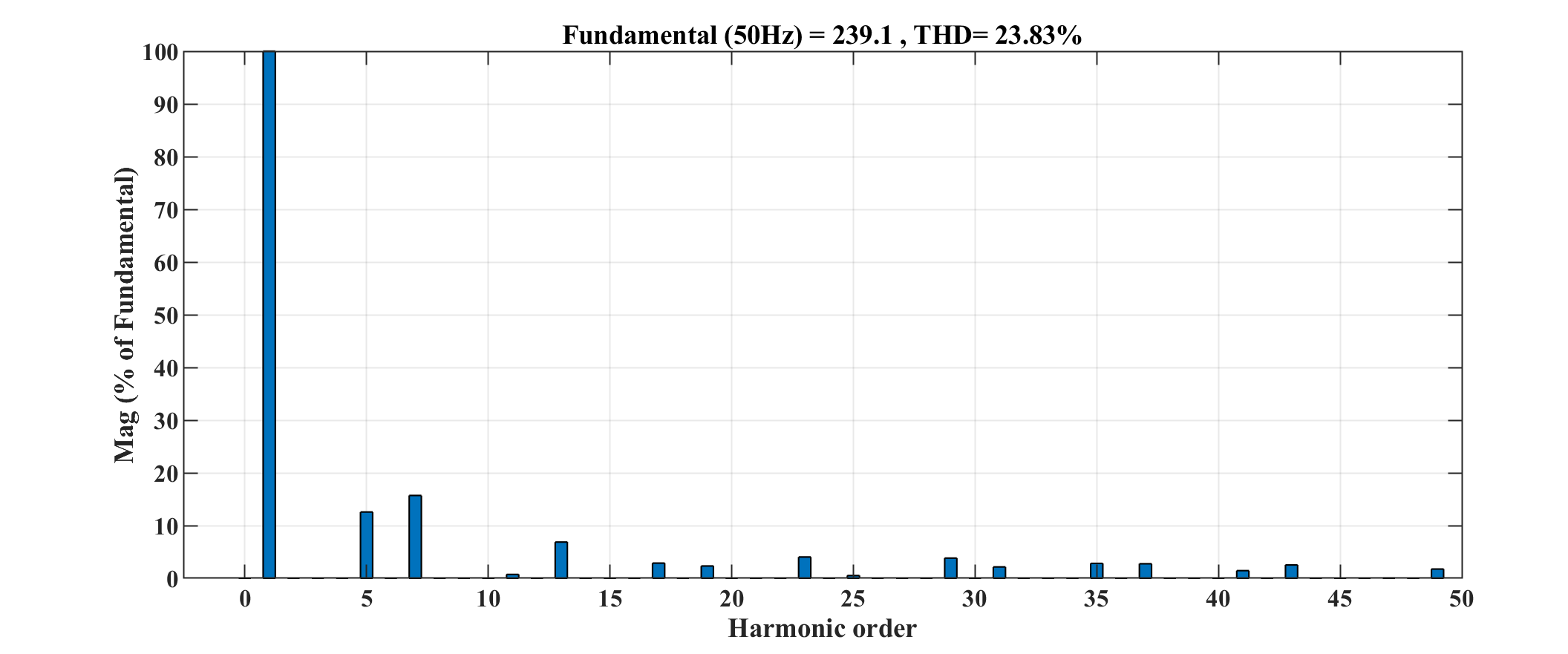} \\
\footnotesize(b)
\end{tabular}
\caption{The proposed system with $V_{o, pu}$=0.6 and $M_{new}$=1. (a) Output voltage waveform. (b) FFT analyze.}
\label{Fig.}
\end{figure}

\begin{table}[]
\caption{Comparing the Harmonic Distortion in Both Conventional and Proposed Methods}
\begin{center}
\begin{tabular}{|c|c|c|}
\hline

\textbf{$V_{o, pu}$} & \textbf{\textit{Conventional Method}}& \textbf{\textit{Proposed Method}} \\
\hline\hline
1& 23.83&23.83  \\
\hline
0.9& 29.91& 23.83 \\
\hline
0.8& 30.9& 23.83 \\
\hline
0.7& 32.69& 23.83 \\
\hline
0.6& 33.10& 23.83 \\
\hline
0.5& 38.71& 23.83 \\
\hline
0.4& 59.30& 23.83 \\
\hline
0.3& 85.91& 23.83 \\
\hline
0.2& 124.19&23.83  \\
\hline
0.1&200.42& 23.83 \\
\hline
\end{tabular}
\label{tab1}
\end{center}
\end{table}
\section{Conclusion}
In this paper, the SHE method's problems have been investigated, and also eliminating the low-order harmonics have been examined. A proposed method for regulating the dc-link voltage of a 5-level CHB inverter is introduced to overcome those challenges. In this method, by adjusting the dc-link voltage, the modulation index is fixed on 1. As a result, the harmonic distortion in the output voltage decreases considerably. Salient examples of this circumstance could be $(V_o)_1$ equal to 120 and 240 volts. In this condition, by comparing the proposed method with the previous one, approximately 60\% and 10\% improvement in the THD amount is observed, respectively. 

\bibliographystyle{IEEEtran}
\bibliography{mybib}

\end{document}